\documentclass[sigconf]{acmart}
\AtBeginDocument{%
  }

\copyrightyear{2025}
\acmYear{2025}
\setcopyright{cc}
\setcctype{by}
\acmConference[CVMP '25]{European Conference on Visual Media Production}{December 3--4, 2025}{London, United Kingdom}
\acmBooktitle{European Conference on Visual Media Production (CVMP '25), December 3--4, 2025, London, United Kingdom}
\acmDOI{10.1145/3756863.3769710}
\acmISBN{979-8-4007-2117-5/2025/12}

\usepackage{multirow}
\usepackage{todonotes}
\usepackage{makecell}

\begin{document}

\title{ReTiDe: Real-Time Denoising for Energy-Efficient Motion Picture Processing with FPGAs}

\author{Changhong Li}
\authornote{Both authors contributed equally to this research.}
\orcid{0009-0003-1340-8611}
\affiliation{%
  \institution{Trinity College Dublin}
  \city{Dublin}
  \country{Ireland}
}
\email{lic9@tcd.ie}

\author{Clement Bled}
\authornotemark[1]
\orcid{0000-0001-5395-5773}
\affiliation{%
  \institution{Trinity College Dublin}
  \city{Dublin}
  \country{Ireland}
}
\email{bledc@tcd.ie}

\author{Rosa Cunningham Fernandez}
\affiliation{%
  \institution{Trinity College Dublin}
  \city{Dublin}
  \country{Ireland}}
\email{cunninr2@tcd.ie}
\orcid{0009-0000-9599-9571}

\author{Shreejith Shanker}
\authornotemark[2]
\authornote{Corresponding author.}
\affiliation{%
  \institution{Trinity College Dublin}
  \city{Dublin}
  \country{Ireland}}
\email{shreejith.shanker@tcd.ie}
\orcid{0000-0002-9717-1804}







\renewcommand{\shortauthors}{Li et al.}

\begin{abstract}
Denoising is a core operation in modern video pipelines. In codecs, in-loop filters suppress sensor noise and quantisation artefacts to improve rate-distortion performance; in cinema post-production, denoisers are used for restoration, grain management, and plate clean-up. However, state-of-the-art deep denoisers are computationally intensive and, at scale, are typically deployed on GPUs, incurring high power and cost for real-time, high-resolution streams. This paper presents Real-Time Denoise (ReTiDe), a hardware-accelerated denoising system that serves inference on data-centre Field Programmable Gate Arrays (FPGAs). A compact convolutional model is quantised (post-training quantisation plus quantisation-aware fine-tuning) to INT8 and compiled for AMD Deep Learning Processor Unit (DPU)-based FPGAs. A client-server integration offloads computation from the host CPU/GPU to a networked FPGA service, while remaining callable from existing workflows, e.g., NUKE, without disrupting artist tooling. On representative benchmarks, ReTiDe delivers 37.71$\times$ Giga Operations Per Second (GOPS) throughput and 5.29$\times$ higher energy efficiency than prior FPGA denoising accelerators, with negligible degradation in Peak Signal-to-Noise Ratio (PSNR)/Structural Similarity Index (SSIM). These results indicate that specialised accelerators can provide practical, scalable denoising for both encoding pipelines and post-production, reducing energy per frame without sacrificing quality or workflow compatibility.
Code is available at https://github.com/RCSL-TCD/ReTiDe.





\end{abstract}



\begin{teaserfigure}
  \includegraphics[width=\textwidth]{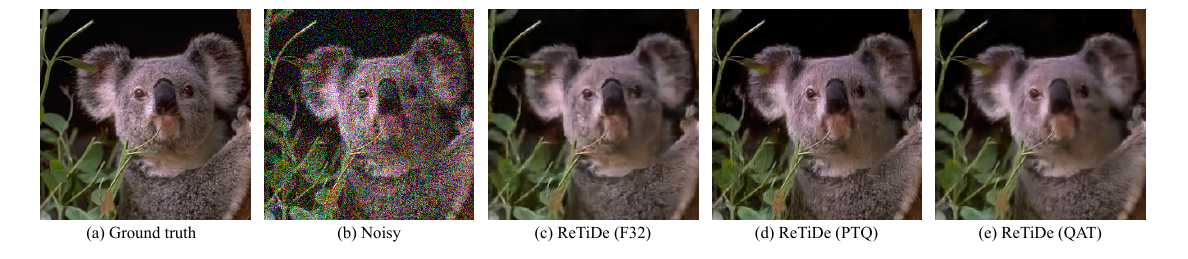}
  \caption{In the BSD100 dataset, when the noise intensity is 45, the ground truth, noisy image and denoising results from the FP32 and the quantised ReTiDe.}
  \label{fig:teaser}
\end{teaserfigure}


\maketitle

\section{Introduction}
Image denoising is a classic ill-posed problem with applications in video compression~\cite{brenig2024higher}, cinematic post-production~\cite{bled2024lightweight}, camera and smartphone imaging pipelines~\cite{wang2024personalized}, and medical imaging~\cite{demir2025diffdenoise}. In video compression, denoising reduces frame entropy, enabling more efficient encoding and lower bandwidth usage. In cinematic post-production, it is essential to clean raw footage before editing and grading. In smartphone imaging, denoising mitigates the higher noise levels inherent to small sensors. In medical imaging, it improves diagnostic clarity by suppressing acquisition noise. 

In recent years, image denoising has shifted from classic methods~\cite{bled2022assessing} based on analytical priors, such as regularisation~\cite{mallat1989theory,combettes2004wavelet,malfait1997wavelet}, wavelet-based denoisers~\cite{mallat1989theory,combettes2004wavelet,malfait1997wavelet}, and nonlocal collaborative filters~\cite{wang2006fast,mahmoudi2005fast,coupe2008optimized, gilboa2009nonlocal, buades2005non}, including the widely used BM3D algorithm~\cite{dabov2007image,dabov2009bm3d}, to data-driven deep learning approaches~\cite{zhang_beyond_2017,zhang2018ffdnet,heinrich_residual_2018,gurrola2021residual,liu_multi-level_2018,wang2020multi,zamir2021multi,qin2020u2,guo_toward_nodate} capable of restoring image details that are otherwise irretrievably lost. While these modern methods generally outperform traditional filters, they often come with substantially higher computational costs, with state-of-the-art transformer architectures~\cite{vaswani2017attention,dosovitskiy2020image,liu2021swin,liang2021swinir} capable of employing hundreds of millions of trainable parameters.

Although research on general-purpose FP32 GPU-accelerated models continues to advance, studies on energy-efficient hardware-accelerated quantised denoising models in this field are limited.
Studies have shown that, with appropriate modifications, quantised neural networks can achieve comparable performance with significantly reduced overhead~\cite{han2015deep}.
FPGAs are an ideal offloading platform for many compute-intensive tasks, including video denoising, due to their latency and power efficiency.
Existing efforts of FPGA-accelerated denoising have mainly focused on optimising classical bilateral filters~\cite{dabhade2017reconfigurable, gabiger2013fpga, wen2024high, spagnolo2024approximate}, whereas research on accelerating quantised deep learning–based denoising models remains largely underexplored.
Recently, studies have demonstrated that deploying deep neural networks such as DnCNN on FPGAs can outperform CPU and GPU implementations in terms of throughput and energy efficiency~\cite{kang2024improvement, tu2024lightweight}.
These approaches rely on specialised quantised convolution Intellectual Property (IP)s based on algorithms such as Winograd fast convolution, which reduce the number of multiplications through matrix transformations and are particularly effective for specific kernel sizes (especially 3×3).
However, such fixed architectures are less suited to advanced models and lack highly parallel pipelined scheduling. Their intermediate results require high-bit storage, which constrains both throughput and energy efficiency.

While post-production workflows now offer FP32 GPU accelerated models, such as The Foundry's catalogue (CATTERY) of deep-learning models~\cite{foundry_cattery_2025}, the advantages of quantised neural network models on FPGAs (in terms of throughput and energy efficiency), particularly for intensive media stream processing, have never been incorporated. To address this, we propose an end-to-end video denoising framework that enables professional users to offload denoising tasks to cloud-based FPGAs via a simple interface, achieving technological decoupling and high-efficiency, low-power video denoising.

The main contributions of this paper are as follows:
\begin{itemize}
    \item We use Post-Training Quantisation (PTQ) to convert FP32 models to INT8 and fine-tune the model with Quantisation-Aware Training (QAT), while maintaining image quality, delivering substantial gains in throughput and energy efficiency with negligible PSNR loss.
    \item We enhance the NUKE--FPGA interface by enabling the plugin to handle larger data chunks, thereby fully utilising PCIe bandwidth. This allows the NUKE plugin to invoke a networked FPGA-based denoising service for real-time use without disrupting the artist workflow.
    \item The quantised ReTiDe-Net achieves denoising quality comparable to popular FP32 baselines. We evaluate its performance across multiple benchmarks, covering both colour and grayscale images.
    \item To the best of our knowledge, ReTiDe is the first open-source, colour, blind, hardware-accelerated denoiser.
\end{itemize}




The remainder of this paper is organised as follows. Section~\ref{background} reviews the background and related work, including image noise characteristics, recent advances in AI-based methods, and prior research on FPGA-based denoising. Section~\ref{methodology} presents our proposed Client-Server denoiser integration framework, detailing the quantised denoising model and deployment strategies. Section~\ref{results} discusses the experimental results, comparing the performance of our model with state-of-the-art methods on both colour and grayscale image denoising tasks. Finally, Section~\ref{conclusion} concludes the paper.

\section{Background and Related Works}\label{background}
\subsection{Noise in Digital Photography}
Although digital camera technology continues to advance, the quantum nature of light imposes a fundamental noise floor in every image. Due to quantum uncertainty, the arrival of photons at discrete photowells follows Poisson statistics, leading to unavoidable fluctuations in the measured photon counts from one photowell to another. This phenomenon, known as photon noise~\cite{beenakker2003quantum, schottky1918spontane, schottky2018spontaneous} (or shot noise), originates outside the sensor’s silicon and defines the minimum achievable noise level in an image. Subsequent amplification of the captured signal within the camera’s image signal processor (ISP) introduces additional noise sources, including dark current noise~\cite{yang_comparative_1999}, fixed-pattern noise~\cite{joseph2001modelling}, and read noise~\cite{liu_photocurrent_2001}.
\subsection{Popular Denoising Algorithms}
Since its introduction in the late 2000s, BM3D~\cite{dabov2007image} has remained a leading classical image denoiser. It performs collaborative filtering of non-local patches in two stages: first, grouping similar patches via block matching, weighting them by similarity, and applying hard thresholding in a transform domain to produce a pilot estimate; second, refining the estimate with a collaborative Wiener (or later, wavelet) filter. Variants of these algorithmic approaches remain in use, for example, Neat Video’s ~\cite{NeatVideo} wavelet-based filter and Wiener-based denoisers in The Foundry’s NUKE~\cite{FoundryNuke} and the AV1~\cite{av1codec} in-loop filter.

The DnCNN~\cite{zhang_beyond_2017} demonstrated clear gains over classical methods by employing an end-to-end 17-layer convolutional network without downsampling. The network predicts the residual image, which is subtracted from the input to produce the denoised output. Its compact size (557k parameters) has contributed to its enduring popularity. Subsequent models built on this foundation, such as IRCNN~\cite{Zhang_2017_CVPR} with dilated convolutions and FFDNet~\cite{zhang2018ffdnet}, which incorporates user-provided noise levels and subsampled inputs.

The encoder–decoder architecture of U-Net~\cite{ronneberger_u-net:_2015} enabled much larger denoising networks, improving quality through multi-scale feature analysis and fusion via skip connections. CBDNet~\cite{guo_toward_nodate} (4M parameters) extends U-Net for blind denoising by incorporating a noise estimation subnetwork alongside the noisy input. MWCNN~\cite{liu_multi-level_2018} and MWRDCNN integrate wavelet transforms into U-Nets, decomposing features into high- and low-frequency components for more efficient processing, and remain among the most competitive non-transformer U-Net variants. Other widely used U-Net denoisers include MPRNet~\cite{zamir2021multi} (20M parameters) and U2Net~\cite{qin2020u2} (44M parameters), reflecting the architecture’s adaptability across diverse denoising tasks.


Most recently, transformer networks~\cite{vaswani2017attention, khan2022transformers, dosovitskiy2020image} have surpassed the performance of CNNs by introducing multi-headed attention layers with global receptive fields, enabling them to capture long-range dependencies beyond local convolutions. The Swin Transformer~\cite{liu2021swin} established a general-purpose backbone by using hierarchical attention layers that progressively downsample features and employ shifted windows to achieve global context, making end-to-end image tokenisation and reconstruction computationally tractable. This architecture has since inspired several denoising networks, including SwinIR~\cite{liang2021swinir}, Uformer~\cite{wang2022uformer}, and Restormer~\cite{zamir2022restormer}.

\subsection{Hardware Accelerated Video Denoising}
Conventional denoisers are typically implemented on CPUs or GPUs. However, their low throughput and limited energy efficiency often become performance bottlenecks in practical applications. Recently, a growing body of research has explored offloading both classical~\cite{dabhade2017reconfigurable, gabiger2013fpga, wen2024high, spagnolo2024approximate} and deep learning-based image denoising algorithms~\cite{kang2024improvement, tu2024lightweight} onto FPGAs to improve performance. Compared to other platforms, FPGAs offer superior throughput, energy efficiency, and reconfigurability, making them an excellent offloading target for such tasks.

Research on optimising hardware-accelerated image denoising algorithms is limited, with even recent work~\cite{yao2022compact,wen2024high,spagnolo2023design,spagnolo2024approximate,xie2024fpga} remaining limited to traditional bilateral filter implementations.
While these examples do not compete with state-of-the-art GPU models in terms of image quality, their simple implementations offer real-time results.
Studies have proposed accelerating the bilateral filtering by simplifying the computation, approximating the filter kernels, and increasing parallelism, thereby enhancing efficiency. In~\cite{dabhade2017reconfigurable}, a constant-time bilateral filtering algorithm using Gaussian polynomial approximation for the spatial kernel was deployed on an FPGA. It enables larger kernels without additional resource overhead. Gabiger et al.~\cite{gabiger2013fpga} achieved pipelined denoising on an FPGA through pixel grouping and clock-level acceleration. Wen et al.~\cite{wen2024high} reduced computational complexity via approximate computing and improved throughput for high-resolution image processing using a data prefetching strategy. Similarly, Fanny et al.~\cite{spagnolo2024approximate} improved energy efficiency using approximation techniques, while preserving real-time and high visual precision.

Storing filter weights in lookup tables (LUTs) allows precomputation on hardware, reducing computational cost. Spagnolo et al.~\cite{spagnolo2023design} approximated the coefficients of both kernels using piecewise functions and encoded them as 7-bit unsigned integers, storing them in reduced-size LUTs. For a $5 \texttimes 5$ kernel, their implementation achieved a maximum operating frequency of 244 MHz and a throughput of 926.8 frames per second. In~\cite{yao2022compact}, Yao et al. proposed a low-cost bilateral filter hardware architecture incorporating a LUT-based divider and a parallelised design, capable of processing 8-megapixel video at 30 frames per second. The implementation of the Gaussian-Adaptive Bilateral Filter (GABF) on FPGA further demonstrates that kernel approximation and pipelining can effectively accelerate the denoising process while maintaining real-time performance \cite{xie2024fpga}.

Recent advances in deep learning for image denoising have shown that its performance has gradually surpassed traditional methods, especially in detail processing, prompting researchers to explore the use of Quantised Neural Networks (QNNs) in denoising accelerators. In recent studies, Tu et al. \cite{tu2024lightweight} achieved 5.39$\times$ and 15.23$\times$ higher energy efficiency compared to GPU and CPU implementations, respectively, by deploying TNet on a ZYNQ FPGA (MZU03A-EG).
Similarly, Kang et al. \cite{kang2024improvement} implemented a quantised DnCNN for denoising, achieving 1.9$\times$ and 26.2$\times$ energy efficiency improvements over GPU and CPU baselines, respectively. These studies highlight the potential of applying QNNs for denoising on FPGAs.
However, their primary focus lies in inference efficiency, lacking a comprehensive denoising performance evaluation. 
TNet-mini provides only limited case studies of denoising on small-scale datasets, 
without presenting statistical evaluations of denoising performance, such as PSNR 
comparisons before and after denoising under different noise environments. 
L-DnCNN reports PSNR results only on the small-scale grayscale datasets 
BSD68 and SET12, and only for limited noise levels (15, 25, and 50), without evaluating 
noise reduction performance on colour images or assessing the method on higher-quality datasets.
Furthermore, these Winograd-based convolutional \cite{kala2019high} accelerators lack flexibility and have poor parallelism, leaving much room for improvement in hardware performance. Additionally, neither implementation has been open-sourced.


To address these limitations, we implement the ReTiDe denoiser, starting from the Cycle-GAN generator~\cite{zhu2017unpaired}, optimising the network for hardware.
Leveraging highly parallel DPU acceleration, the proposed approach further exploits the real-time and energy-efficient characteristics.
%
We purposely train blind models for both colour and grayscale, unlike many existing models, which require user input or the selection of a model trained within a certain noise band.
The quantised denoisers are benchmarked under multiple noise levels for multiple datasets, demonstrating their generalisation capability.
Finally, an end-to-end deployment interface integrated with a server-accelerator architecture, bridging the gap for professional image processing users in effectively harnessing the advantages of hardware acceleration.

\section{Methodology}\label{methodology}

\subsection{Lightweight ReTiDe-Net}
QNNs significantly reduce model size and improve execution efficiency by converting parameters from FP32 to INT8 representations. 
Moreover, mature FP32 operator libraries on GPUs can be mapped to more efficient quantised implementations, where techniques such as operator substitution (e.g., replacing multiplication with shift operations) and operator fusion (e.g., convolution + batch normalisation + activation) transform them into hardware-friendly forms. This greatly enhances hardware efficiency and enables acceleration on fundamental hardware units such as the Digital Signal Processing block (DSP). However, discrepancies between operators and the delayed implementation of quantised operators pose challenges for mapping advanced model structures.

\begin{figure}[htbp!]
    \centering
    \includegraphics[width=0.99\linewidth]{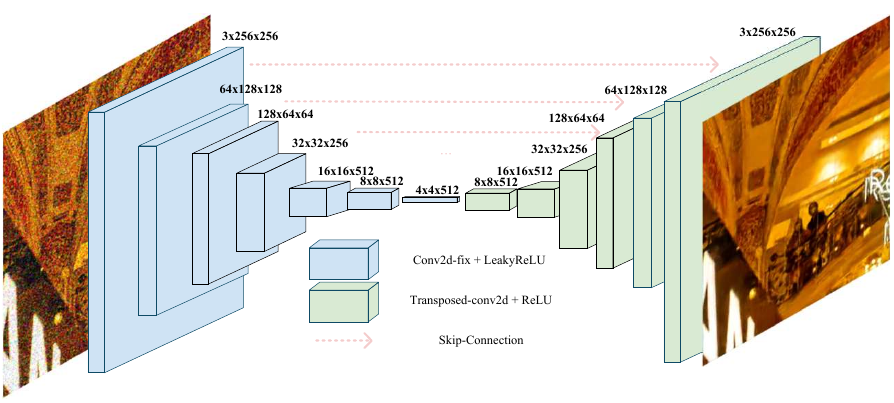}
    \caption{ReTiDe model structure.}
    \label{fig:5}
\end{figure}

Considering this trade-off, we adopt the cGAN backbone~\cite{zhu2017unpaired}, which has been demonstrated as suitable for hardware conversion~\cite{murphy2023custom}, and adapt its U-Net generator while discarding the discriminator. 
The structure of the model is shown in Figure \ref{fig:5}. Our generator is a symmetric encoder–decoder with skip connections between matching stages. In contrast to the original eight-stage design, we employ six downsampling stages, each implemented with a single strided convolution (no bias), paired with a corresponding transposed convolution for upsampling. The innermost block contains both a convolution and a transposed convolution, forming the bottleneck. Skip connections are realised by concatenating encoder features with the decoder output at the same resolution.
To improve hardware efficiency, all downsampling layers employ LeakyReLU activations, 
which prevents gradient vanishing and feature loss that could result from the absence 
of negative values during the downsampling process.
The LeakyReLU coefficient $\alpha=0.1015625$ is approximated as $\tfrac{26}{256}$ on FPGA to enable efficient fixed-point implementation using integer multiplication and bit-shift operations instead of costly floating-point arithmetic.
This enables accurate mapping onto hardware Look Up Table (LUT) operations, avoiding additional 
Block Random Access Memory (BRAM) and DSP resource consumption, thus ensuring both quantisation accuracy and high energy efficiency. 
During upsampling, more regularised feature distributions 
allow the use of ReLU activations, which enhances decoding efficiency.
Normalisation, dropout, and residual connections are omitted to simplify the design and reduce hardware cost, as these modules introduce additional memory access, control logic, and numerical operations that are less hardware-friendly under low-bit quantisation. With this design strategy, we train a grayscale and a colour model and convert both to hardware to compare their performance to existing models.

\subsection{Accelerator Quantisation and Deployment}

\begin{figure}
    \centering
    \includegraphics[width=0.95\linewidth]{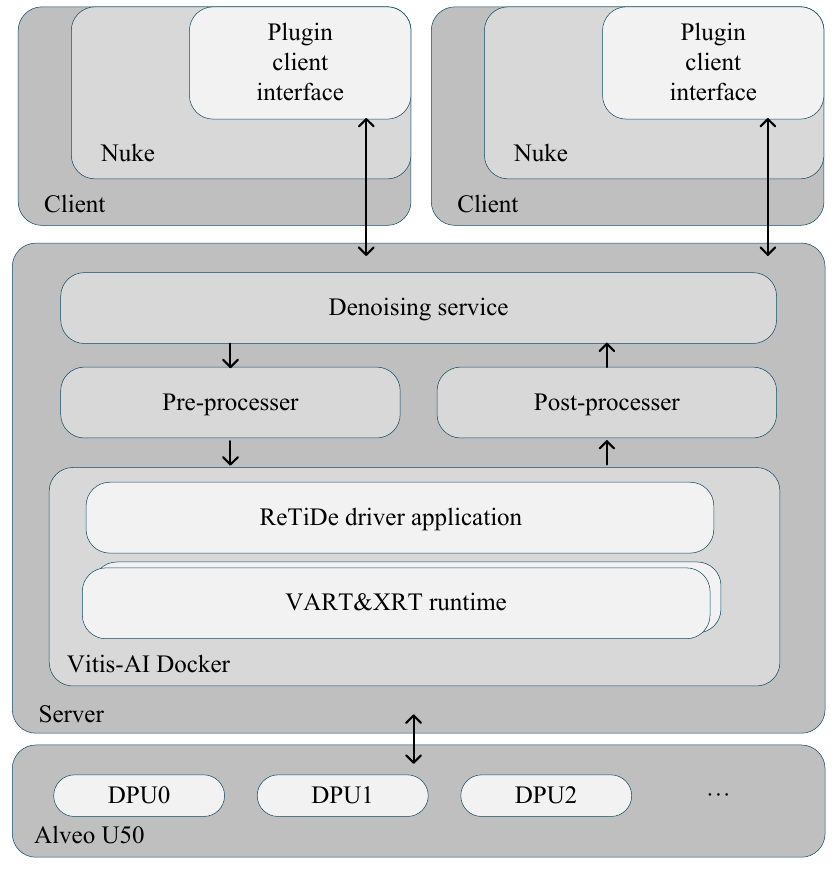}
    \caption{Diagram of the Vitis-NUKE integration.}
    \label{fig:integration}
\end{figure}

The integration workflow of Vitis AI with NUKE is illustrated in Figure~\ref{fig:integration}.
The AMD Vitis AI toolchain provides an end-to-end workflow for deploying quantised neural networks, bridging the gap between machine learning frameworks and FPGA-based deployment.
Moreover, we also provide general client-server interfaces for the integration of other software.
Starting from FP32 model descriptions written in popular frameworks such as TensorFlow and PyTorch, the toolchain performs model quantisation and operator conversion. The converted models can then be accelerated on the Deep Learning Processing Unit (DPU), which is specifically designed for convolutional and matrix-intensive workloads. By mapping computation-intensive kernels directly onto dedicated hardware engines, the toolchain not only reduces CPU overhead but also maximises parallelism and memory bandwidth utilisation. This hardware–software co-design significantly improves inference throughput while simultaneously reducing power consumption.

After training a 32-bit PyTorch FP32 model, Post-Training Quantisation (PTQ) is first applied to convert the original FP32 weights and activations into an 8-bit fixed-point format. PTQ utilises representative data during the calibration phase to capture the distribution characteristics of activations, and then computes scaling factors and zero-point parameters to map continuous FP32 values into a finite integer range. This process substantially reduces model storage overhead and bandwidth requirements, while also allowing the model to better adapt to the hardware logic resources of the FPGA. However, since quantisation inevitably introduces rounding errors and numerical discretisation, model accuracy may be affected. Only calibration is insufficient to adapt the model to quantised inference.

QAT was introduced to mitigate the accuracy degradation caused by quantisation.
In the computational graph, we inserted quantisation stubs and replaced some operators explicitly, resulting in a trainable pseudo-quantised model that emulates quantisation effects during training.
During forward propagation, these nodes approximate computations for weights and activations, while in backwards propagation, gradients are still computed using FP32 parameters.
Through subsequent fine-tune retraining, the pseudo-quantised model adapts to perturbations introduced by quantisation, thereby effectively restoring denoising performance.
A small subset of data is utilised for several forward passes to provide statistical calibration when exporting the quantisation configuration. Finally, the model is quantised and exported for further compilation.

The quantised denoising U-Net model and its quantised weights are loaded into the Vitis AI Docker container for compilation, generating the DPU executable file (xmodel), which can be efficiently mapped onto DPU acceleration tasks on AMD Alveo FPGA accelerator cards. The Alveo platform integrates high-bandwidth memory, fast interfaces, and dedicated management engines, providing strong support for large-scale image processing and inference tasks in data centre environments. To achieve seamless integration with practical rendering workflows, we developed customised runtime drivers that provide an interface between the DPU and the quantised model. This allows the NUKE rendering system to perform multi-threaded parallel invocations through the server-side processing engine, thereby significantly accelerating the inference process while maintaining high accuracy.


\subsection{NUKE Software Interface Integration}


The NUKE Machine Learning Plugin is a dedicated toolkit developed specifically for NUKE, enabling the incorporation of machine learning models into its node-based VFX software environment. To facilitate the integration of the noise reduction function, the noise reduction service interface is encapsulated as a type of NUKE plug-in. Additionally, by modifying the client prior to compilation, the message buffer is extended to accommodate 8K-level data streams. Distributed client hosts can transmit target data to designated servers for real-time rendering by invoking remote deep learning processing services.

\begin{figure}
    \centering
    \includegraphics[width=0.98\linewidth]{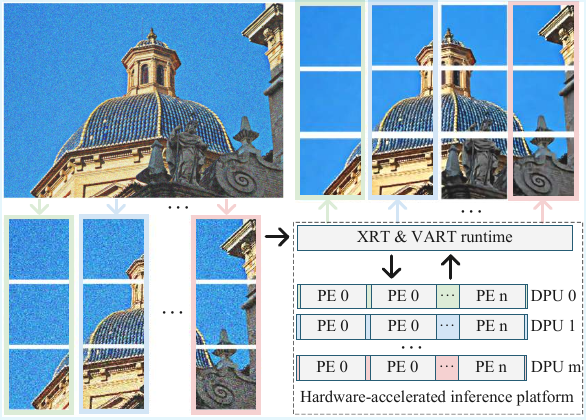}
    \caption{Pre-processing of large input images, parallel hardware-accelerated noise reduction and post-processing.}
    \label{fig:4}
\end{figure}

The server side consists of a host machine equipped with an Alveo U50 server-level FPGA accelerator card. Figure~\ref{fig:4} demonstrates a processing flow from the original noisy image ($\sigma=50$) to the denoised image with segmentation and corresponding parallelisation. Upon receiving denoising requests initiated by remote or local hosts, the incoming image or video stream is first processed by a pre-processor, which segments and batches the media into standardised input formats suitable for the model. This also facilitates parallel processing across multiple threads and DPU units.

The Vitis AI runtime and associated drivers are encapsulated within a standalone Docker image, providing a stable runtime environment and reducing deployment costs. The denoising service invokes the model driver to batch-process the denoising sequences through the VART and XRT runtimes. These sequences are offloaded to the FPGA accelerator via the high-speed PCIe interface in a multithreaded fashion. Subsequently, tasks are distributed in parallel across multiple DPUs optimised for quantised convolution operations, where further parallelisation is achieved through multiple processing elements (PEs). This hierarchical and dedicated parallel processing architecture significantly enhances system throughput and energy efficiency. The post-DPU output is then passed back in reverse order, undergoing post-processing to restore the spatial and temporal sequence before being delivered back to the client.

\begin{figure}
    \centering
    \includegraphics[width=0.98\linewidth]{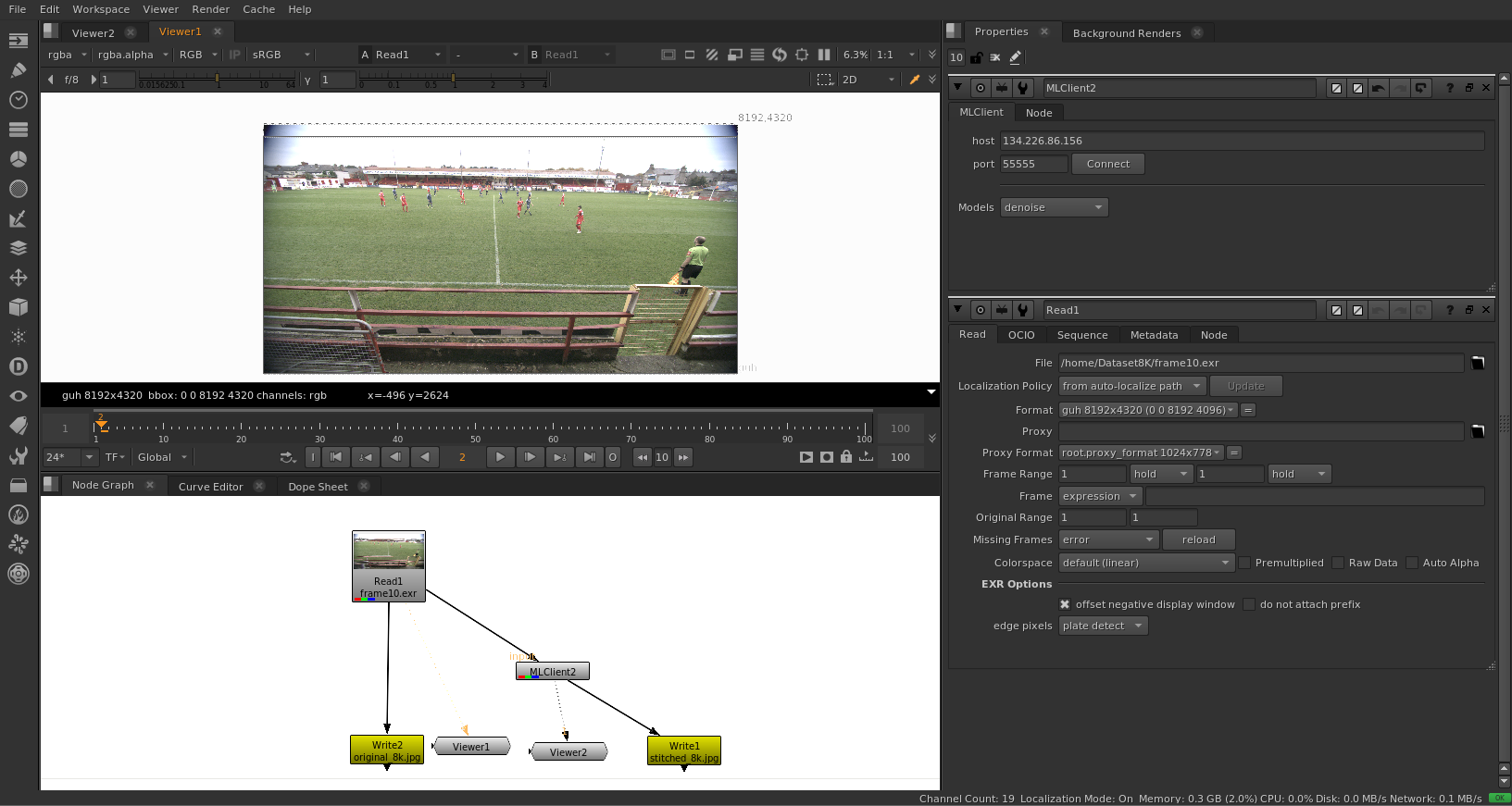}
    \caption{NUKE User Interface.}
    \label{fig:ui}
\end{figure}

The user interface of the client is shown in Figure~\ref{fig:ui}, where an 8K image denoising instance is used for illustration. The upper-left, lower-left, and upper-right sections of the interface respectively display the media to be processed, the node graph corresponding to the workflow, and the connection configuration for the denoising model server. \texttt{Viewer1} and \texttt{Viewer2} nodes are used to preview the input and output of the denoising process, respectively. Additionally, the processed results are stubbed for further access or inspection. Internally, the \texttt{MLClient2} node abstracts the complex hardware-accelerated denoising model into a simple, callable function block that can be easily integrated and triggered within a standard node-based workflow in NUKE. While we integrate the (FPGA) client-side functions through NUKE for our evaluation in this paper, the FPGA client interface is developed to allow seamless integration into other tools and workflows.

\section{Results and Discussion}\label{results}
\subsection{Dataset and Experimental Setups}
The DIV2K~\cite{Timofte_2017_CVPR_Workshops, Agustsson_2017_CVPR_Workshops} and LSDIR~\cite{li2023lsdir} datasets are used for training. Together, they comprise over 85,000 images, although we found that a subset of 4,000 images was sufficient for training the denoiser. These datasets were selected for their high spatial resolution (2K and 4K) and their greater diversity compared to classic computer vision datasets, which typically contain far fewer images at lower resolutions and are often limited to film photography content.


Training is performed on randomly cropped 256×256 patches with a batch size of 64 for 10,000 epochs, using Gaussian noise with noise levels ranging from 0 to 50. For each batch, patches are sampled from the full-resolution images and augmented with random horizontal and vertical flips and random rotations. Optimisation is carried out using AdamW with an initial learning rate of  \(\eta = 10^{-4}\) and weight decay of \(\lambda=10^{-2}\), with a cosine annealing scheduler that decays the learning rate to zero and restarts every 5,000 batches. Model training is conducted in FP32 on an NVIDIA A5000 GPU, using Python 3.13, PyTorch 2.1.1, and CUDA 12.2.

The grayscale versions of datasets BSD68~\cite{martin2001database}, Urban100~\cite{huang2015single} and Set12~\cite{dabov2007image} are used to evaluate the grayscale model and to fairly compare against the existing models~\cite{tu2024lightweight, kang2024improvement}. Colour benchmarks are carried out on BSD100~\cite{martin2001database}.
In the QAT process, we retrained the model for 30 epochs in a quantise-aware manner with a learning rate of $10^{-8}$ to fine-tune and recover the PSNR. Lastly, the Alveo-U50 FPGA was selected as the targeted platform.

To evaluate the performance of our denoising model against existing FPGA-accelerated denoising approaches, we conducted experiments from the following perspectives:  
(1) Comparison of denoising performance in grayscale image denoising with both state-of-the-art FP32 models and FPGA-based deep-learning-accelerated quantised denoising models.  
(2) Comparison of denoising performance in colour image denoising under PTQ and QAT settings with other state-of-the-art FP32 models.  
(3) Comparison of accelerator throughput and energy efficiency with other denoiser models in the literature, where these metrics have been quantified.  
Experimental results demonstrate that our integrated solution not only introduces a high-throughput and high-energy-efficiency quantised denoising scheme into the NUKE workflow, but also achieves denoising performance comparable to, or surpassing, that of state-of-the-art FP32 and quantised models.

\subsection{Grayscale Denoising Evaluation}

Grayscale denoising involves only single-channel input, which implies a reduction in the amount of input data available to the model. This naturally leads to differences in performance compared to colour denoising. To benchmark our model against existing denoising accelerators, we first evaluated denoising performance on grayscale images. Experiments were conducted on the classical BSD68 dataset and the higher-quality URBAN100 grayscale dataset, comparing our model against existing FP32 models, particularly FPGA-implemented quantised denoising models. For PSNR baseline tests in grayscale denoising, and to remain consistent with prior benchmarks, three noise levels with standard deviations of 15, 25, and 50 were employed. The experimental results are summarised in Table~\ref{tab:grayscale_denoising}, with the PTQ and QAT models referred to as ReTiDe\,(P) and ReTiDe\,(Q) respectively.  From the data, it can be observed that our quantised model achieves denoising performance close to that of 32-bit FP32 models under 8-bit quantisation. Moreover, in high-noise scenarios on the BSD68 dataset, our model outperforms the existing denoising accelerator L-DnCNN.

\begin{table}[htbp]
\centering
\caption{PSNR ($dB$) comparison of various algorithms for grayscale image denoising.}
\label{tab:grayscale_denoising}
\begin{tabular}{@{ }llcccccc@{}}
\toprule
& \multirow{2}{*}{Method} & \multicolumn{3}{c}{BSD68 (gray)} & \multicolumn{3}{c}{URBAN100 (gray)} \\
\cmidrule(lr){3-5} \cmidrule(lr){6-8}
 & & 15 & 25 & 50 & 15 & 25 & 50 \\
\midrule
\multirow{6}{*}{\rotatebox[origin=c]{90}{FP32}} & BM3D     & 30.95 & 25.32 & 24.89 & 31.91 & 29.06 & 24.45 \\
& FFDNet     & 31.45 & 28.96 & 25.16 & 33.76 & 31.41 & 28.09 \\
& IRCNN     & 31.46 & 28.79 & 25.11 & 33.08 & 29.62 & 24.53 \\
& DnCNN-20 & 31.60 & 29.14 & 26.20 & 33.76 & 30.19 & 19.34 \\
& SwinIR     & 31.76 & 29.10 & 25.40 & 33.44 & 30.43 & 25.47 \\
& \textbf{ReTiDe}   & 31.48 & 29.09 & 26.20 & 33.25 & 30.60 & 26.55 \\
\midrule
\multirow{3}{*}{\rotatebox[origin=c]{90}{QNNs}} & L-DnCNN & 31.44 & 29.01 & 26.08 & - & - & - \\
& \textbf{ReTiDe (P)}   & 29.92 & 28.35 & 26.73 & 29.92 & 28.35 & 25.52 \\
& \textbf{ReTiDe (Q)}   & 30.94 & 29.23 & 26.73 & 30.20 & 28.46 & 25.61 \\
\bottomrule
\end{tabular}
\end{table}

\begin{figure}[htbp!]
    \centering
    \includegraphics[width=0.95\linewidth]{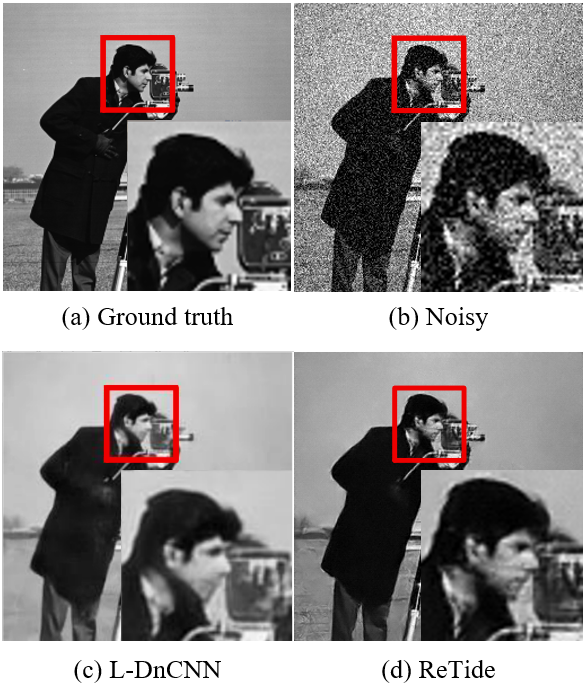}
    \caption{Comparison of output results with other quantised image denoising models under the noise level of 35.}
    \label{fig:set12_comparison}
\end{figure}

\begin{table*}{}
\centering
\caption{PSNR ($dB$) and SSIM comparison of popular denoisers for colour denoising on BSD100.}
\label{tab:psnr_bsd100}
\begin{tabular}{@{ }llcccccccccccc@{}}
\toprule
&\multirow{4}{*}{Method} & \multirow{4}{*}{\shortstack{Blind/\\Nonblind}} & \multicolumn{10}{@{}c@{}}{BSD100 (colour)} \\
\cmidrule(lr){4-13}
&& & \multicolumn{2}{c}{5} & \multicolumn{2}{c}{15} & \multicolumn{2}{c}{25} & \multicolumn{2}{c}{35} & \multicolumn{2}{@{}c@{}}{45} \\
\cmidrule(lr){4-5} \cmidrule(lr){6-7} \cmidrule(lr){8-9} \cmidrule(lr){10-11} \cmidrule(lr){12-13}
&& & PSNR & SSIM & PSNR & SSIM & PSNR & SSIM & PSNR & SSIM & PSNR & SSIM \\
\midrule
\multirow{5}{*}{\rotatebox[origin=c]{90}{FP32}} & BM3D         & Nonblind & 39.85 & 0.98    & 33.17 & 0.9223 & 30.16 & 0.8598 & 28.17 & 0.8007 & 26.62 & 0.747  \\
& DnCNN        & Blind & 39.72 & 0.9728  & 33.46 & 0.9245 & 30.56 & 0.8711 & 28.62 & 0.8217 & 27.11 & 0.7766 \\
& FFDNet       & Nonblind    & 39.84 & 0.9788  & 33.65 & 0.9265 & 31.00 & 0.8772 & 29.37 & 0.8337 & 28.23 & 0.7958 \\
& IRCNN        & Nonblind    & 39.95 & 0.9789  & 33.41 & 0.9234 & 30.45 & 0.8678 & 28.43 & 0.8114 & 26.87 & 0.7578 \\
&\textbf{ReTiDe} & Blind & 39.46 & 0.9761  & 33.27 & 0.9205 & 30.65 & 0.8682 & 29.03 & 0.8224 & 27.89 & 0.7826 \\
\midrule
\multirow{2}{*}{\rotatebox[origin=c]{90}{QNNs}} & \textbf{ReTiDe (P)} & Blind & 32.94 & 0.8943       & 30.38 & 0.8414      & 28.95 & 0.8149      & 27.96 & 0.7941      & 27.06 & 0.7713      \\
& \textbf{ReTiDe (Q)} & Blind & 33.22 & 0.9425       & 31.03 & 0.9008      & 29.37 & 0.8576      & 28.14 & 0.8164      & 27.14 & 0.7811      \\
\bottomrule
\end{tabular}
\end{table*}

\begin{figure*}
    \centering
    \includegraphics[width=0.8\linewidth]{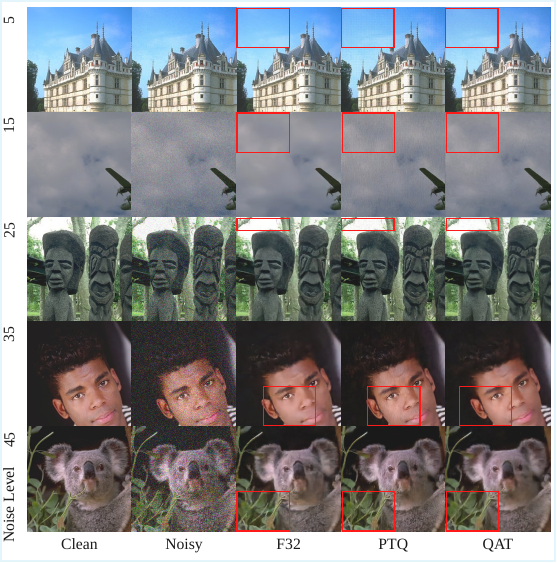}
    \caption{Denoising results of FP32 and quantised ReTiDe denoising models at different noise levels.}
    \label{fig:denoising_results}
\end{figure*}

While PSNR reflects overall denoising performance, it does not capture all aspects of denoising; human visual perception and preservation of image details are equally important. The Set12 dataset is a classical baseline dataset; however, due to its early origin, the ground truth itself contains deviations from clean images. As shown in Figure~\ref{fig:set12_comparison}(a), the ground truth images contain significant noise. Therefore, this dataset is primarily used for a detailed comparison of denoising results with other accelerators. Figure~\ref{fig:set12_comparison}(b) shows images corrupted with Gaussian noise at a level of 35, and detailed denoising result comparisons are made between L-DnCNN (c) and our quantised model (d). The results indicate that our model significantly outperforms L-DnCNN in detail preservation, particularly in facial features such as the mouth and hair contours, where the features are retained with higher accuracy and realism.

The superior detail-preserving capability of ReTiDe compared to L-DnCNN can be attributed to its encoder-decoder architecture with skip connections.
This enables the model to integrate high-level semantic information while preserving fine details. In contrast, DnCNN relies solely on deep convolutional stacking, which tends to smooth fine details across multiple abstraction layers.





\subsection{Colour Denoising Evaluation}

In addition to PSNR performance comparisons and detailed evaluations against existing baseline denoising accelerators, we further validate our quantised model ReTiDe on the colour dataset BSD100, which is closer to real-world application scenarios. We conduct baseline tests under Gaussian noise with finer-grained standard deviations of 5, 15, 25, 35, and 45, comparing our method with other state-of-the-art FP32 models. To better capture perceptual quality from a user perspective, we also introduce the SSIM as an evaluation metric. The experimental results are summarised in Table~\ref{tab:psnr_bsd100}. Notably, under blind denoising (denoising without input of the type or intensity of the noise), our model achieves performance comparable to or exceeding that of classical and advanced FP32 models, while QAT further improves both PSNR and SSIM metrics. In high-noise conditions, such as a noise level of 45, our 8-bit quantised model is only 1.09 dB and 0.0147 away from the best 32-bit FP32 model in terms of PSNR and SSIM, respectively. In contrast, under low-noise conditions, quantisation error becomes the dominant factor instead of noise, resulting in relatively larger performance gaps compared to FP32 models. We anticipate that hybrid-precision quantisation could mitigate this issue in future work, though such an exploration is beyond the scope of this paper.

To further assess the denoising detail preservation of our quantised model on colour images, we present results in Figure~\ref{fig:denoising_results}, showing clean, noisy, and denoised images under different noise levels for the FP32 model, the PTQ model, and the QAT model. We highlight regions with noticeable differences. Our observations show that quantisation errors and quantisation-induced noise introduced by PTQ can lead to inferior denoising performance, especially in relatively smooth background regions. However, QAT’s quantisation-aware fine-tuning effectively alleviates this problem, achieving denoising quality nearly identical to that of the FP32 model. These denoising examples further demonstrate the robustness of our model on colour images across multiple noise levels.


\subsection{Deployment Performance}
We deployed the quantised model in the form of an \texttt{xmodel} on a server equipped with an Alveo U50 FPGA, while the corresponding FP32 models were deployed on a client equipped with an NVIDIA A4000 GPU and an Intel(R) Core(TM) Ultra 7 265K CPU. Batch denoising tasks were invoked via software Application Programming Interfaces (APIs) to evaluate the runtime throughput and energy efficiency of the quantised model. A one-minute warm-up inference was first performed to stabilise device operation, after which throughput was computed based on FPS and runtime performance. GPU and CPU power consumption were measured using the \texttt{powerstat} tool, FPGA power was measured with \texttt{xbutil} tool, by recording the average power difference between the IDLE state and active inference, and this value was adopted as the energy efficiency metric. The experimental results are summarised in Table~\ref{tab:platform_performance}.

\begin{table}[htbp!]
\centering
\caption{Performance comparison across platforms: Frequency, Throughput, Power, and Energy Efficiency.}
\label{tab:platform_performance}
\begin{tabular}{@{}lcccc@{}}
\toprule
Method & \makecell{Platform} & \makecell{Thr. \\ (GOPS)} & \makecell{Power \\ (W)} & \makecell{Energy\\ Eff. \\ (GOPS/W)} \\
\midrule
L-DnCNN      & I7-7700HQ CPU & 29.5   & 45   & 0.66 \\
TNet-mini        & I5-12400F CPU & 164.3   & 65 & 2.53 \\
ReTiDe        & U7-265K CPU  & 770.2    & 42.1   & 18.30 \\
\midrule

L-DnCNN       & RTX 1070 GPU & 1066.7    & 115   & 9.28 \\
TNet-mini       & RTX 2080Ti GPU & 1785.7   & 250  & 7.14 \\
ReTiDe        & A4000 GPU & 8,285.5   & 236.3  & 35.06 \\
\midrule

L-DnCNN& MZU03A-EG FPGA & 41.8     & 2.4  & 17.18 \\
TNet-mini & MZU03A-EG FPGA  & 99.3     & 2.6  & 38.51 \\
\textbf{ReTiDe}  & Alveo U50 FPGA   & 3,746.1  & 18.4     & 203.59 \\
\bottomrule
\end{tabular}
\end{table}

During this evaluation, FP32 models were run on the CPU and GPU, while quantised INT8 models were run on the FPGA. The results demonstrate that our FPGA-based denoising inference achieves a throughput 37.71$\times$ higher than other FPGA-based baseline deep learning denoisers, reaching 3,746.09 GOPS. Compared to the 0.033s inference time of L-DnCNN~\cite{kang2024improvement}, our model takes only 0.004s. This is attributed to our quantised inference optimisations tailored for multi-DPU architectures and multi-threaded parallel execution. Although TNet-mini~\cite{tu2024lightweight} and L-DnCNN~\cite{kang2024improvement} achieve strong performance through the use of the Winograd algorithm and lightweight CNN structures, server-level FPGAs typically provide more efficient architectures, including high-bandwidth memory (HBM) and customizable DPUs.
Although the PCI Gen 3x4 interface on the xDMA for U50 limited its overall bandwidth, Vitis-AI offers comprehensive system-level optimisations, such as operator fusion, layer-wise quantisation, and efficient memory scheduling, which significantly improve energy efficiency in practical deployment scenarios.

For energy efficiency, our approach surpassed baseline denoising accelerators by 5.29$\times$, reaching 203.59 GOPS/W. These results indicate that our quantised denoising accelerator and its end-to-end deployment scheme deliver a significant advantage in processing large-scale denoising tasks with high energy efficiency compared to conventional hardware platforms, including CPUs, GPUs, and existing FPGA-based neural network denoising accelerators.

\section{Conclusion}\label{conclusion}
This work proposes a hardware-accelerated image denoising solution integrated into professional media processing software. The end-to-end framework offloads computationally intensive denoising tasks to a server-level FPGA DPU acceleration platform. Through multi-threading and multi-PE parallel quantised acceleration, the framework significantly improves real-time denoising throughput and energy efficiency. This approach bridges the gap between the advantages of state-of-the-art quantised denoising models in terms of real-time performance and energy efficiency, and the demands of professional image media processing software for handling computationally intensive media denoising tasks.
For denoising performance, the proposed solution achieves results close to advanced FP32 models for both colour and grayscale images. Compared with existing FPGA-based deep learning denoising accelerators, it achieves 5.29$\times$ and 37.71$\times$ improvements in energy efficiency and throughput, respectively. This offers a new pathway for accelerating and offloading professional image processing algorithms.
Future work will focus on further enhancing denoising detail performance using mixed-precision quantisation on more advanced models, exploring model sparsification to achieve even greater energy efficiency, conducting systematic model ablation studies to gain deeper insights into design trade-offs, and incorporating full pipeline overhead estimation and optimisation alongside real-world image noise to improve the practical applicability of the proposed solution.


\begin{acks}
This work was funded by the Horizon CL4 2022 - EU Project Emerald – 101119800.
\end{acks}


\bibliographystyle{ACM-Reference-Format}

\bibliography{reference}

\end{document}